\documentclass[aps,prl,showpacs,twocolumn,groupedaddress]{revtex4}
\usepackage{graphicx,amsmath,amssymb}
\begin{document}
\newtheorem{Theorem}{Theorem}
\newtheorem{Lemma}{Lemma}

\title{Linearity and Quantum Adiabatic Theorem}

\author{Zhaohui Wei}
\email{weich03@mails.tsinghua.edu.cn }
\author{Mingsheng Ying}
\email{yingmsh@mail.tsinghua.edu.cn }

\affiliation{ State Key Laboratory of Intelligent Technology and
Systems, Department of Computer Science and Technology, Tsinghua
University, Beijing, China, 100084}

\begin{abstract}

We show that in a quantum adiabatic evolution, even though the
adiabatic approximation is valid, the total phase of the final state
indicated by the adiabatic theorem may evidently differ from the
actual total phase. This invalidates the application of the
linearity and the adiabatic approximation simultaneously. Besides,
based on this observation we point out the mistake in the
traditional proof for the adiabatic theorem. This mistake is the
root of the troubles that the adiabatic theorem has met. We also
show that a similar mistake remains in some recent modifications of
the traditional adiabatic theorem attempting to eliminate the
troubles.

\end{abstract}

\maketitle

The adiabatic theorem \cite{BF28,LIS55} is an important result in
quantum theory, which has a long history. Since discovered, it has
been widely applied in various areas of physics such as nuclear
physics, quantum field theory and so on. This theorem states that if
the the state of a quantum system is governed by a slowly changing
Hamiltonian, and if the initial state of this system is one of the
eigenstates of the initial Hamiltonian, then the state of the system
at any time will be the corresponding eigenstate of the Hamiltonian
of that time up to a multiplicative phase factor. About two decades
ago, when studying circular adiabatic evolution, Berry discovered a
new phenomenon called geometric phase \cite{Berry84}, which has been
generalized to many other cases \cite{AA87,SB88,Uhl86,ES00}.
Besides, since quantum information science bloomed, quantum
adiabatic theorem has been employed essentially in two very
important schemes of quantum computation, namely geometric quantum
computation \cite{ZR99,JVEC00} and quantum adiabatic computation
\cite{FGGS00}.

On the other hand, adiabatic theorem has been reexamined in the
recent years. Some authors \cite{MS04,TSKO05,TSKFO05} noted that one
must be careful in employing the adiabatic theorem. Concretely,
Marzlin and Sanders demonstrated that a perfunctory application of
the theorem may lead to an inconsistency \cite{MS04}. Furthermore,
Tong et al. pointed out that the root of the inconsistency is that
the traditional adiabatic conditions are not sufficient to guarantee
the validity of adiabatic approximation \cite{TSKO05}. To eliminate
the inconsistency, some modified adiabatic conditions have been
proposed, such as those in Refs.\cite{YZZG05,DMN05,COM06}.

In this letter, we propose a new approach to revisit quantum
adiabatic theorem. We show that, in an adiabatic evolution, even if
the adiabatic approximation is valid under some conditions  , the
total phase indicated by the adiabatic theorem may differ evidently
from the actual value. Here, ``the validity of the adiabatic
approximation'' means that the state of the system equals
approximatively to some eigenstate of the Hamiltonian up to a phase
factor all the time. For convenience, we call the phase of this
phase factor total phase. Generally, the global phase of a quantum
state is not important from an observational point of view. However,
the global phase can be changed to a relative phase when we combine
the adiabatic approximation and the linearity, the well-known
property of quantum mechanics. Different relative phases may give
rise to physically observable differences, so our result points out
that the linearity of adiabatic approximation may fail. Therefore,
errors in the approximation of phase may lead to severe problems
when linearity is also used. On the other hand, by reexamining the
adiabatic theorem based on this observation, we reveal the mistake
in its traditional proof. It is this mistake that makes the
traditional adiabatic conditions not sufficient to guarantee the
validity of adiabatic approximation. At last, we show that there are
similar problems with some modified adiabatic conditions
\cite{YZZG05,DMN05}.

For convenience of the later presentation, let us recall the quantum
adiabatic theorem and its traditional proof.

Suppose we have a quantum system and the evolution of its state
$|\psi(t)\rangle(t_0\leq t)$ is governed by a time-dependent
Hamiltonian $H(t)$. Suppose the initial state
$|\psi(t_0)\rangle=|E_{k}(t_0)\rangle$ is an eigenstate of the
initial Hamiltonian $H(t_0)$. In the instantaneous eigenbasis
${|E_n(t)\rangle}$ of $H(t)$, the state can be expressed as
\begin{equation}
|\psi(t)\rangle=\sum_{n}\psi_{n}(t)e^{-i\int E_n}|E_n(t)\rangle,
\end{equation}
where $\int E_n=\int^{t}_{t_0}E_n(t')dt'.$ Substituting Eq.(1) into
the Schr\"{o}dinger equation
\begin{equation}
i\frac{d}{dt}|\psi(t)\rangle=H(t)|\psi(t)\rangle,
\end{equation}
we obtain the following differential equation:
\begin{equation}
i\dot{\psi_{n}}=-i\sum_{m}e^{i\int(E_n-E_m)}\psi_{m}\langle
E_n|\dot{E_m}\rangle.
\end{equation}
If it holds that
\begin{equation}
|\langle E_n|\dot{E_m}\rangle|\ll |E_n-E_m|, \ \ n\neq m,
\end{equation}
the phase factor $\exp{(i\int(E_n-E_m))}$ in Eq.(3) will be a rapid
oscillation. This can make transitions to other levels negligible,
and results in
\begin{equation}
|\psi(t)\rangle\approx e^{-i\int E_k}e^{-\int \langle
E_k|\dot{E_k}\rangle}|E_{k}(t)\rangle.
\end{equation}
Here we call Eq.(4) the traditional adiabatic conditions. The
standard adiabatic theorem states that, if these conditions hold, we
have approximation Eq.(5).

One can also prove this theorem as follows. Suppose
\begin{equation}
|e_i(t)\rangle=e^{-\int \langle E_i|\dot{E_i})}|E_i(t)\rangle, \ \
i=1,...,N.
\end{equation}
In the new instantaneous eigenbasis ${|e_n(t)\rangle}$ of $H(t)$,
the state can be expressed as ($\psi_n$'s are different from those
of the above)
\begin{equation}
|\psi(t)\rangle=\sum_{n}\psi_{n}(t)e^{-i\int E_n}|e_n(t)\rangle.
\end{equation}
Substituting Eq.(7) into the Schr\"{o}dinger equation gives
\begin{equation}
i\dot{\psi_{n}}=-i\sum_{m\neq n}e^{i\int(E_n-E_m)}\psi_{m}\langle
e_n|\dot{e_m}\rangle.
\end{equation}
Similarly, if
\begin{equation}
|\langle e_n|\dot{e_m}\rangle|=|\langle E_n|\dot{E_m}\rangle|\ll
|E_n-E_m|, \ \ n\neq m,
\end{equation}
we know that the integral of the rhs (right hand side) of Eq.(8) is
negligible, due to which we get (let $n=k$)
\begin{equation}
\psi_{k}(t)-\psi_{k}(t_0)\approx 0.
\end{equation}
Note that
\begin{equation}
\psi_{k}(t_0)=1.
\end{equation}
We finally get
\begin{equation}
|\psi(t)\rangle\approx e^{-i\int E_k}e^{-\int \langle
E_k|\dot{E_k})}|E_k(t)\rangle,
\end{equation}
which finishes the proof of the traditional adiabatic theorem once
more.

From Eq.(5) we know that after an adiabatic evolution, the state of
the system has an approximate total phase. Now we show that there
may be a significant gap between this approximate total phase and
the corresponding exact total phase. To be concrete, we use a simple
but interesting example to show this point (See also
\cite{TSKFO05,TSKO05}).

Let us consider a spin-half particle in a rotating magnetic field.
The Hamiltonian of the system can be written as
\begin{equation}
H(t)=-\frac{\omega_0}{2}\begin{pmatrix}
\cos{\theta} & \sin{\theta}e^{-i\omega t} \\
\sin{\theta}e^{-i\omega t} & -\cos{\theta}\\
\end{pmatrix},
\end{equation}
where $\omega_0$ is a time-independent parameter defined by the
magnetic moment of the spin and the intensity of external magnetic
field, $\omega$ is the rotating frequency of the magnetic field. The
instantaneous eigenvalues and eigenstates of $H(t)$ are
\begin{equation}
E_1=\frac{\omega_0}{2}, \ \ E_2=-\frac{\omega_0}{2},
\end{equation}
\begin{equation}
|E_1(t)\rangle=\begin{pmatrix}
e^{-i\omega t/2}\sin{\frac{\theta}{2}}\\
-e^{i\omega t/2}\cos{\frac{\theta}{2}}\\
\end{pmatrix},
\end{equation}
\begin{equation}
|E_2(t)\rangle=\begin{pmatrix}
e^{-i\omega t/2}\cos{\frac{\theta}{2}}\\
e^{i\omega t/2}\sin{\frac{\theta}{2}}\\
\end{pmatrix},
\end{equation}
respectively. The adiabatic conditions Eq.(4) are satisfied as long
as
\begin{equation}
\omega_0\gg \omega\sin{\theta}.
\end{equation}

Suppose that the system is initially in the state $|E_1(0)\rangle$.
At time $t$, according to the adiabatic theorem, the system will be
in the instantaneous eigenstate $|E_1(t)\rangle$ up to a phase
factor. This approximate total phase $\gamma(\tau)$ can be
calculated by Eq.(5),
\begin{equation}
\gamma(\tau)=-\int E_0+i  \int \langle E_0|\dot{E_0}).
\end{equation}
Substituting Eqs.(14)-(16) into Eq.(18) gives
\begin{equation}
\gamma(\tau)=-\frac{\tau}{2}(\omega_0+\omega\cos{\theta}).
\end{equation}
On the other hand, we can evaluate this total phase almost without
any approximation. Let us express the exact state of the system as
\begin{equation}
|\psi(t)\rangle=a(t)|E_1(t)\rangle+b(t)|E_2(t)\rangle.
\end{equation}
By solving the Schr\"{o}dinger equation it can be checked that
\begin{equation}
a(t)=\cos{\frac{\bar{\omega}t}{2}}-i\frac{\omega_0+\omega\cos{\theta}}{\bar{\omega}}\sin{\frac{\bar{\omega}t}{2}},
\end{equation}
\begin{equation}
b(t)=i\frac{\omega\sin{\theta}}{\bar{\omega}}\sin{\frac{\bar{\omega}t}{2}},
\end{equation}
where
$\bar{\omega}=\sqrt{\omega^2+\omega_0^2+2\omega\omega_0\cos{\theta}}$.

When the traditional adiabatic condition Eq.(17) is satisfied,
$b(t)\rightarrow 0$, so $|\psi(t)\rangle\approx a(t)|E_1(t)\rangle$.
Then we can regard the exact total phase as
\begin{eqnarray*}
\aligned \hat{\gamma}(\tau)=&\arg{(\cos{\frac{\bar{\omega}\tau}{2}}-i\frac{\omega_0+\omega\cos{\theta}}{\bar{\omega}}\sin{\frac{\bar{\omega}\tau}{2}})}\\
\approx &-\frac{\bar{\omega}\tau}{2}.
\endaligned
\end{eqnarray*}
We finally get
\begin{eqnarray*}
\aligned \gamma(\tau)-\hat{\gamma}(\tau)\approx &\frac{\bar{\omega}\tau}{2}-\frac{\tau}{2}(\omega_0+\omega\cos{\theta})\\
= &\frac{1}{2}\cdot
\frac{\omega^2\sin^2{\theta}}{\bar{\omega}+(\omega_0+\omega\cos{\theta})}\tau.
\endaligned
\end{eqnarray*}
That is to say, the difference between $\gamma(\tau)$ and
$\hat{\gamma}(\tau)$ increases approximately linearly with the
running time $\tau$ (See FIG. 1 for details, where
$d(\tau)=\hat{\gamma}(\tau)-\gamma(\tau)$ and $t_0 = 2\pi/\omega$ is
the period of the Hamiltonian $H(t)$). According to Fig. 1, we know
that the difference becomes remarkable when the running time is very
long.

\begin{figure}[htb]
\centerline{\includegraphics[angle=0,width=3.3in]{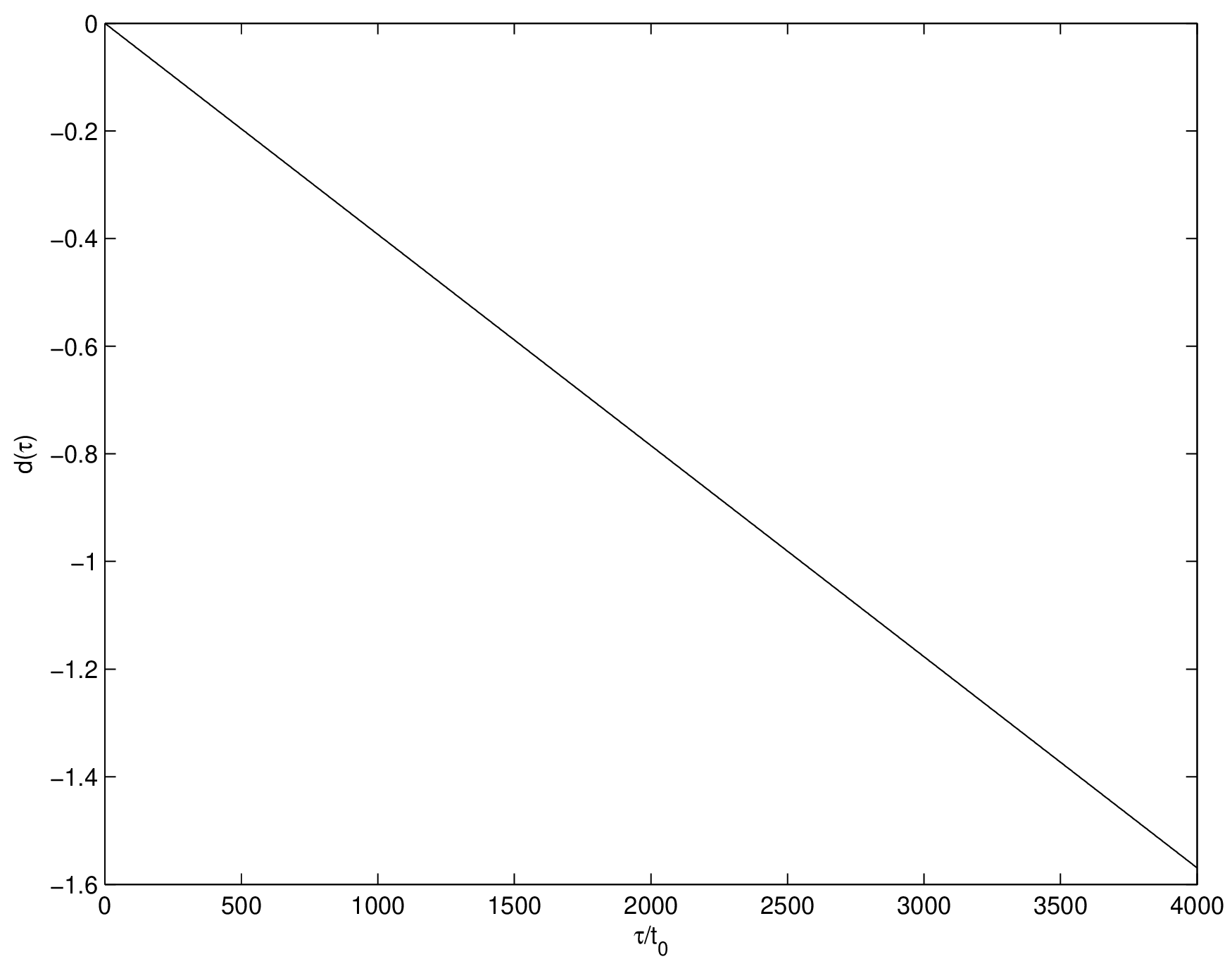}}
\caption{Computer simulated result of the difference between the
approximate total phase and the exact total phase, where
$\omega=0.01$, $\omega_0=10$,$\theta=\pi/6$.} \label{fig:gaps_Id}
\end{figure}

Thus we have shown the total phase given by the traditional
adiabatic theorem may have a remarkable error if the running time is
very long. As a result, we must be careful if we combine the
linearity of quantum mechanics and adiabatic approximation.
Concretely, we can use the previous example to illustrate this
point. Suppose the initial state of the system is
\begin{equation}
|\psi(0)\rangle=\frac{1}{\sqrt{2}}|E_1(0)\rangle+\frac{1}{\sqrt{2}}|E_2(0)\rangle.
\end{equation}
Note that no matter which eigenstate the initial state is, the
traditional adiabatic conditions are the same, because
\begin{equation}
|\langle E_1|\dot{E_2}\rangle|=|\langle E_2|\dot{E_1}\rangle|.
\end{equation}
If the linearity of adiabatic approximation is valid, and if the
common adiabatic condition
\begin{equation}
|\langle E_1|\dot{E_2}\rangle|\ll |E_1-E_2|.
\end{equation}
is satisfied, it can be checked that the final state of the system
will be
\begin{equation}
|\psi(\tau)\rangle\approx\frac{1}{\sqrt{2}}e^{-\alpha}|E_1(\tau)\rangle+\frac{1}{\sqrt{2}}e^{\alpha}|E_2(\tau)\rangle,
\end{equation}
where, $\alpha=i\int E_1+\int \langle E_1|\dot{E_1}\rangle.$

However, based on Fig. 1, we know that there must be some $\tau$
such that the exact state of the system
\begin{equation}
|\psi'(\tau)\rangle\approx\frac{1}{\sqrt{2}}e^{-\alpha-\frac{\pi}{2}i}|E_1(\tau)\rangle+\frac{1}{\sqrt{2}}e^{\alpha+\frac{\pi}{2}i}|E_2(\tau)\rangle.
\end{equation}
Note that $\langle\psi(\tau)|\psi'(\tau)\rangle\approx 0$. That is
to say, the final state indicated by the linearity of adiabatic
approximation is far from the exact final state. Thus, in this case
the linearity of adiabatic approximation fails.

We have mentioned that the traditional adiabatic theorem has been
criticized recently. Thus it is necessary to turn back to its proof
and find out what is the problem. We will scrutinize the proof with
the previous example in mind.

According to the proof, if the traditional adiabatic conditions are
fulfilled the integral of the rhs of Eq.(8) will be small no matter
how long the running time $\tau$ is. However, the example above tell
us that this is not true. When gap between these two total phase is
remarkable, this integral cannot be neglected. Thus there must be
something wrong with the proof. Now we still use the previous
example to find out the mistake. We think of the mistake as the root
of the problem of the traditional adiabatic theorem.

Comparing Eq.(7) and Eq.(20), we have
\begin{equation}
\psi_2(t)=i\frac{\omega\sin{\theta}}{\bar{\omega}}\sin{\frac{\bar{\omega}t}{2}}e^{-i\frac{\omega_0
t}{2}}e^{-i\frac{\omega\cos{\theta}}{2}t}.
\end{equation}
At this time, the rhs of Eq.(8) is ($n=1$)
\begin{equation}
e^{i\int (E_1-E_2)}\psi_2\langle
e_1|\dot{e_2}\rangle=-\frac{(\omega\sin{\theta})^2}{\bar{\omega}}\sin{\frac{\bar{\omega}t}{2}}e^{i\frac{\omega_0+\omega\cos{\theta}}{2}t}.
\end{equation}
The proof for the traditional adiabatic theorem at the beginning of
this letter tells us that if the adiabatic condition
\begin{equation}
\frac{1}{2}\omega\sin{\theta}=|\langle e_1|\dot{e_2}\rangle|\ll
|E_1-E_2|=\omega_0
\end{equation}
is satisfied, the integral of the rhs of Eq.(29) can be negligible.
Now we show that this is not true. In fact,
\begin{eqnarray*}
\aligned
Im(\int_{0}^{\tau}\sin{\frac{\bar{\omega}t}{2}}e^{i\frac{\omega^{*}}{2}t}dt)=&\int_{0}^{\tau}\sin{\frac{\bar{\omega}t}{2}}\sin{\frac{\omega^{*}}{2}t}dt\\
=&\frac{1}{2}\int_{0}^{\tau}\cos{\frac{\bar{\omega}-\omega^{*}}{2}t}dt\\
-&\frac{1}{2}\int_{0}^{\tau}\cos{\frac{\bar{\omega}+\omega^{*}}{2}t}dt\\
=&\frac{1}{\bar{\omega}-\omega^{*}}\sin{\frac{\bar{\omega}-\omega^{*}}{2}\tau}\\
-&\frac{1}{\bar{\omega}+\omega^{*}}\sin{\frac{\bar{\omega}+\omega^{*}}{2}\tau},
\endaligned
\end{eqnarray*}
where $\omega^{*}=\omega_0+\omega\cos{\theta}$. Suppose $\omega_0\gg
\omega\sin{\theta}$, it can be checked that
\begin{equation}
Im(\int_{0}^{\tau} e^{i\int (E_1-E_2)}\psi_2\langle
e_1|\dot{e_2}\rangle dt)\approx
-2\sin{\frac{\bar{\omega}-\omega^{*}}{2}\tau}.
\end{equation}
At the same time,
\begin{equation}
Re(\int_{0}^{\tau} e^{i\int (E_1-E_2)}\psi_2\langle
e_1|\dot{e_2}\rangle dt)\approx
2(\cos{\frac{\bar{\omega}-\omega^{*}}{2}\tau}-1).
\end{equation}
It is obvious that $\sin{\frac{\bar{\omega}-\omega^{*}}{2}\tau}$ may
be far from $0$ when $\tau$ is big. So the integral of the rhs of
Eq.(29) cannot be negligible for all $\tau$. Thus a mistake in the
proof for the traditional adiabatic theorem is found. When the
traditional adiabatic conditions are satisfied, though the phase
factor in Eq.(3) or Eq.(8) provides a rapid oscillation, the
coefficients $\psi_i$ may do the same thing. This makes the integral
that has been ignored not neglectable. This is the reason why the
traditional adiabatic conditions are not sufficient to guarantee the
validity of adiabatic approximation.

After the inconsistency was found, several attempts have been made
to repair the traditional adiabatic conditions, for example
Ref.\cite{YZZG05} and Ref.\cite{DMN05}. If we reexamine these new
conditions, we will find that there are similar problems with the
proofs for them. According to these two papers, in Eq.(8) we should
consider not only the absolute value of $\langle
e_n|\dot{e_m}\rangle$ but also its phase. Then when we integrate the
rhs of Eq.(8) the integrand has a new phase factor. For the example
we have considered Eq.(8) will be
\begin{equation}
i\dot{\psi_1}=-i\frac{1}{2}\omega\sin{\theta}\cdot\psi_2\cdot
e^{i(\omega_0+\omega\cos{\theta})t}.
\end{equation}
Now the modified adiabatic condition is
\begin{equation}
\omega_0+\omega\cos{\theta}\gg \omega\sin{\theta}.
\end{equation}
However, we can see that in these two proofs the effects of
$\psi_n$'s were not considered. In fact, as we have pointed out,
$\psi_n$'s result in the consequence that the integral we have
ignored cannot be neglected even if the modified adiabatic
conditions are fulfilled.

Before concluding, we should emphasize that we cannot say quantum
adiabatic theorem is invalid, though the traditional adiabatic
conditions are insufficient. For an example of rigorous and
sufficient but a litter complicated adiabatic conditions, we can see
Ref.\cite{AR04}. This result states that in an adiabatic evolution
if the path along which the Hamiltonian of the system varies is
fixed (when the running time changes, the path does not), the
adiabatic approximation will be valid as long as the evolution is
slow enough. Thus this result guarantees the validity of quantum
adiabatic computation \cite{FGGS00}.

Another point that should be stressed is that in the above example
the running time $\tau$ that makes the adiabatic traditional proof
fail is very long (See FIG. 1). When $\tau$ is not very long, the
negligence in the proof is not fatal, and the adiabatic
approximation is very good. Furthermore, perhaps in many cases, the
phases of the $\psi_n$'s that we ignore is trivial (change slowly),
then the adiabatic conditions are sufficient. As a consequence, the
traditional adiabatic conditions are free of problems most of the
time, but not always.

In conclusion, we have shown that in a quantum adiabatic evolution
the approximate total phase of the final state proposed by Berry may
differ evidently from the corresponding exact total phase when the
running time is long enough. Because of this difference, we have to
be very careful if we use both the linearity and quantum adiabatic
approximation, especially when the running time of the adiabatic
evolution is very long.

On the other hand, based on this difference, we have reexamined the
traditional proof of the quantum adiabatic theorem. We show that
there is an evident problem in the proof. We think this problem is
the origin of the troubles the the traditional adiabatic theorem has
met. This problem also exists in the proofs for some modifications
of the traditional adiabatic conditions. In fact, these proofs all
are based on the fact the integrals of rapid oscillations vanish. It
seems that our discussion demonstrates that this is not a good way
to lead to rigorous adiabatic conditions. Because we have known that
$\psi_n$'s may have a considerable contribution to the integral.
However, without solving the Schr\"{o}dinger equation we cannot know
anything about $\psi_n$'s.

We acknowledge D. M. Tong and Peter Marzlin for valuable comments
and suggestions. We also thank C. P. Sun and the colleagues in the
Quantum Computation and Information Research Group for helpful
discussions. This work was partly supported by the National Nature
Science Foundation of China (Grant Nos. 60503001, 60321002, and
60305005).

\end{document}